# Emergent Behavior in Strongly Correlated Electron Systems

David Pines, Santa Fe Institute and Physics Department, U C Davis, and UIUC

## Abstract

I describe early work on strongly correlated electron systems [SCES] from the perspective of a theoretical physicist who, while a participant in their reductionist top-down beginnings, is now part of the paradigm change to a bottom-up "emergent" approach with its focus on using phenomenology to find the organizing principles responsible for their emergent behavior disclosed by experiment---and only then constructing microscopic models that incorporate these. After considering the organizing principles responsible for the emergence of plasmons, quasiparticles, and conventional superconductivity in SCES, I consider their application to three of SCES's sister systems, the helium liquids, nuclei, and the nuclear matter found in neutron stars. I note some recent applications of the random phase approximation and examine briefly the role that paradigm change is playing in two central problems in our field: understanding the emergence and subsequent behavior of heavy electrons in Kondo lattice materials; and finding the mechanism for the unconventional superconductivity found in heavy electron, organic, cuprate, and iron-based materials.

## Introduction

A few years ago, in a unit I wrote for the Annenberg Learner Project's "Physics for the Twenty-first Century" [1], I described the early work on strongly correlated electron systems [SCES] and their sister systems in quantum matter, the helium liquids, nuclei, and the celestial nuclear liquids found in pulsars. I did so through an emergent lens in which the focus was on the emergent behavior these display. Because my audience was high school teachers and beginning undergraduates, I included a minimum of equations and references. In the present perspective, I revisit this territory from the perspective of one who is part of the paradigm change from reductionism, based on a "first principles" Hamiltonian description of electron interaction, to an "emergent" paradigm that focuses on using phenomenology to identify the organizing principles in SCES that are responsible for their emergent behavior before constructing microscopic models that incorporate these, models that need not be based on a first principles Hamiltonian description [2] [3].

Following a review of the electron interaction landscape in 1948, the year I started research on electron interactions in classical and quantum plasmas at Princeton under the direction of David Bohm, I describe the contents of my 1950 Ph.D thesis, "The Role of Plasma Oscillations in Electron Interactions", and review what has been called the Bohm-Pines quartet [4], the four papers bearing the title of "A Collective Description of Electron Interactions" [5], of which the first three of which were written with Bohm, and the fourth by me. In these papers on classical and quantum plasmas, the random phase approximation [RPA] played a central role in making possible a microscopic account of the hallmarks of their emergent behavior to which electron interaction gives rise: collective plasma oscillations and quasi-electrons that interact via a screened Coulomb interaction.

I then discuss the extension of the collective description to include ionic motion that enabled John Bardeen and me to propose that our net effective attractive phonon-induced interaction between electrons near the Fermi surface could be the organizing principle for conventional superconductivity [6], and so set the stage for BCS [7]. Following a review of how plasmons emerged as a well-defined elementary excitation in nearly all solids during the period 1954-56 [8], I discuss briefly a twenty-first century application for plasmons, *plasmonics*, that has emerged as a significant sub-field of nano-electronics [9]. I next review the way in which phenomenology has made it possible to identify the effective interactions responsible for the emergence and subsequent interactions between the elementary excitations in the helium liquids, with results that are in remarkably good agreement with experiment [10], and consider how collective concepts in condensed matter have led to our understanding of the giant dipole resonance in nuclei [11], nuclear superfluidity, [12], and the most abundant form of superfluid matter found in the universe, the celestial nuclear superfluids found in the crust and core of pulsars [13]. After a brief review of the ways the RPA continues to impact our understanding of emergent behavior in many-body physics, I examine briefly some of the ways in which paradigm change is influencing research on two major challenges to the SCES community: understanding the emergence and subsequent behavior of heavy electrons in Kondo lattice materials; and establishing the mechanism for the unconventional superconductivity found in cuprates, heavy electron, and organic superconductors.

**The Electron Interaction Landscape in 1948**

Landau famously said, "You cannot repeal Coulomb's law", but he said nothing about ignoring it; apart from a few papers by Wigner and Bardeen in the 1930's, that was pretty much the situation in the solid state community before 1948. Although the cohesive energy of metals, as calculated in the best microscopic description available, the Hartree-Fock approximation, was off by almost an order of magnitude because of its failure to include correlations between electrons of opposite spin, there was no systematic microscopic way to improve upon that result. The best effort to deal with the *correlation energy* of metallic electrons, the difference between their true ground state energy and that calculated in the Hartree-Fock approximation, was made by Wigner, who arrived at a phenomenological expression for it by carrying out a bold interpolation of the correlation energy he had obtained at densities that differed by two orders of magnitude: a variational calculation for an electron gas that he expected might be valid at a density some ten times higher than seen in metals; and the electron solid he predicted would be found at densities some ten times lower[14]. When applied to the cohesive energy of the alkali metals, the resulting expression turned out to be remarkably successful.

Matters were not better when one considered the behavior of individual electrons in metals. Although the independent particle model of electrons in metals in which electron interactions were ignored seem to work well, Bardeen's attempt to go beyond it, by calculating the corrections to single electron energies brought about by the Coulomb interaction between electrons, had led to divergent results and a T/lnT term in the specific heat that was not seen experimentally [15]. A decade later, Wohlfarth proposed a phenomenological approach to deal with this logarithmic divergence — assume that the effective interaction was in fact screened with a range of the order of the inter-electron spacing [16]. This yielded reasonable results, but if one then tried to use this interaction to calculate the exchange energy, one found that the cohesive energy of the alkali metals is positive, rather than being negative and large.

**The Role of Plasma Oscillations in Electron Interactions**

I started my Ph.D thesis research at Princeton University in the fall of 1948. My thesis advisor and mentor, David Bohm, suggested I explore whether the main effect of the Coulomb interaction between electrons in a quantum plasma, the interacting electrons moving in a background of uniform positive charge that Conyers Herring called jellium, might be to bring about collective plasma oscillations, as was the case in the classical plasmas on which he was an expert. In my 1950 Ph.D thesis, I developed a detailed physical picture of the origin of collective behavior in plasmas and of the modifications in the motion of single electrons as they responded to one another, and found a way to incorporate these basic concepts into a Hamiltonian formulation of the many-electron problem, so that one could move easily from a classical to a quantum description, and, a la Schwinger's approach to QED, use canonical transformations to decouple the collective plasma modes from the individual electron behavior.

I first studied transverse collective modes and magnetic electron interactions. A virtue of the Hamiltonian approach is that it provides a unified picture in which one can see directly both the influence of the electron plasma on the transverse collective modes and the influence of those collective modes on the motion of the electrons in the plasma. For the transverse plasma oscillations-electromagnetic waves modified by their coupling to the electron plasma—decoupling the collective modes from the single particle motion in the plasma turned out to be a comparatively easy task, provided I made what I called in my thesis the plasma approximation, and was later called the *random phase approximation [RPA]:* for a given momentum transfer, q, keep only the qth component of the relevant interactions. After a fair amount of trial and error, I found the decoupling canonical transformation from the starting Hamiltonian for light waves coupled to the electron plasma to a Hamiltonian in which the modified fields and the modified electrons were no longer coupled.

The decoupled electromagnetic waves were the renormalized transverse plasma oscillations—electromagnetic waves whose dispersion relation, as modified by their coupling to the electron plasma, took the form,

$$\omega_q^2 = c^2 q^2 + \omega_p^2 \qquad [1]$$

where $\omega_p$, the electron plasma frequency, is given by

$$\omega_p = [4\pi n e^2/m]^{1/2} \quad [2]$$

and n is the density of electrons of mass m. The decoupled electrons possessed an effective interaction that was a screened version of the weak, but long-range, magnetic interaction between a pair of free electrons of velocity $v_i$ and $v_j$ separated by a distance $r_{ij}$ that is induced by their coupling to the transverse electromagnetic field [the Biot Savart law for two electrons]

,
$$-[e^2/r_{ij}][\mathbf{v}_i \cdot \mathbf{v}_j/c^2]\exp -[\omega_p/c] r_{ij} \quad [3]$$

The physics behind the screening is straightforward; in the RPA, electron j sees both the bare electron i and the magnetic current its motion induces in the plasma; the latter acts to screen its field, so their effective interaction is given by Eq.[3]. It was not difficult to show that the above results hold whether one is dealing with a classical or quantum plasma.

I developed an analogous approach to the collective longitudinal modes of a classical plasma by working in a longitudinal gauge in which, for an isolated pair of electrons, it is the vector potential that gives rise to their long range Coulomb interaction when one imposes a subsidiary condition on it. What emerged after a series of canonical transformations was the longitudinal counterpart of the above results: a new Hamiltonian in which the collective and individual electron contributions were decoupled. The collective term described longitudinal plasma oscillations whose frequency is given by their classical dispersion relation,

$$1 = [4\pi e^2/m] \Sigma_i 1/([\omega - q \cdot v_i]^2 \quad [4]$$

where $\mathbf{v}_i$ is an electron velocity; the second term described what we would now call quasiparticles—electrons plus their co-moving screening clouds--whose effective interaction is

$$[e^2/r]\exp -[\omega_p/<v>] r, \quad [5]$$

while the subsidiary condition that connects the collective field variables to the density fluctuations was shown to vanish within the RPA.

I had shown that the long range part of the Coulomb interaction gave rise only to the collective plasma oscillations; once this was accounted for, the remaining particle interactions are screened, with an effective range ~ $<v>/\omega_p$, the Debye screening length, where $<v>$ is the average particle velocity at a given temperature. Just as for the transverse case, the screening occurs in the RPA because a given electron j sees both the bare electron i and the density fluctuation its motion induces in the plasma, with the latter screening the field of electron i.

However, unlike the transverse case, problems appeared to arise when one goes to the quantum plasma, because it appeared difficult to show that when the various decoupling transformations were carried out, the subsidiary condition could be satisfied at a quantum level. As a result, I did not present a quantum Hamiltonian approach in my thesis. I gave instead a self-consistent field derivation [within the RPA] of the dispersion relation for quantized plasma oscillations; for a given wavevector q, this was given by

$$1 = [4\pi e^2/m] \Sigma_k 1/([\omega - q \cdot v_k]^2 - [\hbar^2 q^4/4m^2]) ; \quad [6]$$

the quantum corrections to the classical dispersion relation are seen to be small. I argued that in the quantum domain one would expect to find results similar to those I had obtained classically; the long range of the Coulomb interaction would give rise to quantized plasma oscillations and once these were taken into account, one would find electrons interacting by a quantum analogue of the classical screened Coulomb interaction,

$$[e^2/r]\exp -[\omega_p/v_f] r \quad [7]$$

whose range is $\sim v_f/\omega_p$, where $v_f$ is the Fermi velocity. This would provide a microscopic justification for Wolhlfarth's phenomenological calculation of the electronic specific heat.

In the remainder of my thesis, I described work in progress on following the behavior of the density fluctuations of the quantum plasma through an approach that makes clear the way in which the RPA leads to the quantum dispersion relation, and suggested a simple appealing ground state wave function that takes into account the long range correlations by writing it as a product of a wavefunction representing long wavelength quantized plasma oscillations in their ground state and the usual Slater determinant for free electrons, $\Phi$,

$$\Psi = (\exp[-\Sigma_{k<kc} \rho_k \rho_{-k}/\alpha_k]) \Phi \quad [8]$$

where $\alpha_k$ is a normalization factor.

**A Collective Description of Electron Interactions: 1951-53**

The first paper in this series, I. Magnetic Interactions, was a straightforward presentation of the results described above for the transverse collective modes and magnetic electron interactions. II. Collective vs Independent Particle Aspects of the Interactions, was based in part on my thesis, but also reflected research carried out with Bohm in the year after I received my Ph.D. In it we described the organizing principles responsible for what we would now call emergent behavior in a classical plasma, by focusing on its density fluctuations, $\rho_q$, whose equation of motion is given by

$$d^2\rho_q/dt^2 = -\omega_p^2 \rho_q - \Sigma_i(\mathbf{q}\cdot\mathbf{v}_i)^2 \exp[-i\mathbf{q}\cdot\mathbf{x}_i] - \Sigma_{i,k}(4\pi e^2/mk^2)\mathbf{q}\cdot\mathbf{k}[\exp(i[\mathbf{k}-\mathbf{q}]\cdot\mathbf{x}_i)]\exp(-i\mathbf{k}\cdot\mathbf{x}_j) \quad [9]$$

where **k** is not =**q** in the last term on the rhs of Eq.(9). We argued that to the extent that the phases in the last term in [9] were random, [which we expected to be the case for most physical situations] it would be small and could be neglected, and presented a quantitative justification of this *random phase approximation*. From [9] it is clear that at long wavelengths the density fluctuations then oscillate at the plasma frequency, with corrections of order q² arising from the second "kinetic" term on its rhs; for q> [$\omega_p$/<v>], the kinetic term will dominate, so the density fluctuations at these wavevectors would describe weakly correlated independent particles. We built on this simple picture to show how the density fluctuations could be split into two parts: a collective component that carried out plasma oscillations; and a "quasi-electron" component—electrons plus their associated screening clouds—whose effective interaction would be short range and weak and given approximately by Eq.[5].

Importantly, we used the correspondence principle to give a quantitative explanation of the characteristic energy losses that had been measured in experiments on the scattering of kilovolt electrons by thin metallic films of by Ruthemann [17] and Lang [18], whose work had been brought to our attention by Conyers Herring. We showed that an electron moving faster than the mean thermal velocity of electrons in a plasma would excite plasma waves, and that the characteristic energy losses seen for Al and Be were almost certainly their quantized valence electron plasma oscillations, while our calculated mean free path for their excitation was close to that seen experimentally. For Al the calculated quantum of energy loss, $\hbar\omega_p$, is 14.7 ev, while the measured characteristic energy loss was ~ 15.9 ev; for Be, the corresponding numbers are 18.8 ev and 19.0 ev. From Lang's estimate of the thickness of his films, we concluded that the mean free path for emission of a plasma quantum would be < 185A, while our calculated value for this quantity was ~150A.

It thus turned out that collective oscillations of valence electrons in metals at nearly the free electron plasma frequency, the central topic of my Ph.D. thesis, were being discovered experimentally at about the same time as these were being proposed theoretically, while the stage was set for the further experiments on the scattering of fast electrons by thin solid films. As discussed below, these led, within a few years, to the unambiguous identification of *plasmons*, the quantized plasma oscillations of free valence electrons, as the principal source of the measured characteristic energy losses for almost all solids that were investigated.

The third paper in the series, III. Coulomb Interactions in a Degenerate Electron Gas, brought to fruition the research problem David Bohm had given me in 1948. It was written entirely by correspondence, as by the time we started work on it, Bohm was in Sao Paolo as his first stop in his McCarthy-era forced scientific exile from the US. In it we presented a longitudinal version of the quantum Hamiltonian approach that had been developed for magnetic interactions in Paper I. We introduced field variables to describe the long wavelength collective modes and subsidiary conditions that related these to the density fluctuations, and carried out a series of canonical transformations to decouple these from single electron motion. On making the RPA, we arrived at a comparatively simple final Hamiltonian,

$$H = H_{coll} - 2\pi n e^2 \sum_{k<k_c} [1/k^2] + H_{qp} \quad [10]$$

where $H_{coll}$ describes n' long-wave length [$k<k_c$] collective plasma oscillations whose frequency is given by their quantum dispersion relation, Eq.[6], the second term is the Coulomb self energy of the charge distribution that is now described by the collective modes, and $H_{qp}$ describes what we would now call quasi-particles, whose effective mass was $m^*=m/[1—n'/3n]$ and whose interaction was given by $H_{rp} + H_{sr}$. $H_{rp}$ was a low momentum transfer [$k<k_c$] weak attractive velocity-dependent interaction coming from the coupling of electrons to plasma waves that we argued could be neglected for most purposes, while

$$H_{sr} = 2\pi e^2 \Sigma_{k>k_c} [1/k^2] \qquad [11]$$

was the short-range part of the Coulomb interaction that within the RPA would not be affected by the plasma modes [and vice-versa].

We noted that the upper limit, $k_c$, to the wavevectors for which it is appropriate to introduce plasma modes was of order the quantum Debye length, $\omega_p/v_f$, and that it might be useful to determine it by minimizing the total ground state energy; we proposed that the influence of $H_{sr}$ on the ground state energy and single particle excitations could be treated by perturbation theory. We further showed that after the decoupling canonical transformations, the subsidiary condition involved only particle coordinates. We gave a number of arguments that suggested its influence on the physical properties of the system would be small and discussed as well the connection of our approach to Tomonaga's collective approach to fermion systems [19], showing that it led to his results for a one-dimensional system.

The fourth paper, IV. Electron Interaction in Metals, demonstrated that the "Collective Description" worked well in practice. The correlation energy I calculated agreed well with experiment for the alkali metals, while the corrections to single particle properties such as the electronic specific heat were shown to be sufficiently small that these could be neglected to a first approximation, so the collective description provided the first microscopic justification for the success of the independent electron model of metallic behavior. In a subsequent paper [20] I showed that electron interaction did have appreciable consequences for one "single particle" property, the Pauli spin susceptibility, and that my results for this quantity agreed well with its first direct measurement, for Lithium and Sodium, by Schumacher, Carver, and Slichter [21].

An invitation to give the opening lecture of the 1954 Solvay Congress on the topic of "The Collective Description of Electron Interaction in Metals" gave me an opportunity to present the above results to a distinguished international audience, and their response was most encouraging [22]. One of the unique features of a Solvay Congress is that there is ample time for discussion after each report. In my case the participants in the discussion made significant additional contributions to the topic. Neville Mott gave a beautifully simple explanation [based on the f-sum rule] of why the measured characteristic energy loss in many insulators and most metals is so often close to the free valence electron plasma quantum energy, and he, and later in the discussion, Herbert Frohlich, connected the plasma frequency to the frequency at which the dielectric function vanishes; Harry Jones discussed of the x-ray band width of polyvalent metals, and John Van Vleck suggested that my results for the ground state energy might be obtained with a variational wave function of the form, Eq.[8], proposed in my Ph. D thesis, an approach that was pursued by Gaskell and was later suggested independently by Feynman in a letter to me.

In the years that immediately followed there were a number of further developments in our understanding of emergent behavior in quantum plasmas:

*Gell-Mann and Brueckner developed a diagrammatic approach to the calculation of the ground state energy]; on summing all the terms in the perturbation theory that appeared in the RPA, they obtained a rigorous result that was shown by them to be valid in the weak coupling limit, $r_s$ <1, corresponding to densities large compared to those found in metals [23], while Gell-Mann [24] calculated the specific heat in this same weak coupling limit.

*It became clear that while there were many different ways to derive our results on plasmons and screening, the simplest way to do so was to introduce the frequency and wavevector dependent dielectric function; this approach, which had been suggested by Mott and Frohlich in Brussels, was first carried out by Lindhard [25] and Hubbard [26].

*Improved calculations of the correlation energy at metallic densities that agreed well with experiment were carried out by Hubbard [26] and Nozieres and me [27]; in common with the earlier approaches, these were based on using the RPA to deal with only the long-range part of the Coulomb interaction.

*It turned out that one could ignore altogether the subsidiary conditions that had been of such concern to Bohm and me in developing our collective description. Because the added terms introduced to obtain a Hamiltonian description of the collective modes represent an external probe of the system, one could construct a quite general argument that the associated subsidiary conditions could always be safely neglected. I gave demonstrating this as a problem for students in a graduate course I was teaching in 1960 and it may be found in the text based on it [28].

*It was subsequently recognized that the long wavelength emergent behavior of electron liquids and gases, screening and collective oscillation, is "protected" [2] in that whether one deals with a classical plasma or a quantum electron liquid, the exact screening length is $s/\omega_p$, where s is the adiabatic sound velocity, while $\omega_p$ is the exact frequency at long wavelengths of their plasma oscillations [29].

*The application of polarization potential theory [see below] to the electron liquid made it possible to develop a simple physical picture of the local field corrections that describe the influence of the short-range interactions between electrons on the dielectric function, and calculate its consequences for a number of metallic properties [30].

Looking back, since the coupling constant, $r_s$, the interelectron spacing measured in unit of the Bohr radius, for electrons at metallic densities is large [$r_s$, ~ 3.2 for Li, 4 for Na] my early calculations of the correlation energy and the conduction electron paramagnetic susceptibility for these materials represented the first successful microscopic calculations of the properties of a strongly correlated electron system, the electron liquid.

**Setting the stage for BCS**

After finishing work on the papers described above, I was eager to see whether I could extend our collective description to include ionic motion and electron-phonon interactions. Doing so seemed important because while the experiments of the Maxwell and Serin groups [31] on the influence of isotopic mass on the superconducting transition temperature had made it clear that phonons played an important role in superconductivity, Bardeen and Frohlich had independently found that including the coupling to phonons in the electron self energy did not yield superconductivity. Frohlich had then suggested that perhaps it was the phonon-induced interaction between electrons that led to superconductivity [32], but his proposal had received little attention because he had not included the expected much larger role of the screened Coulomb interaction between electrons.

Tor Staver, who succeeded me as Bohm's graduate student, had started work on the problem with Bohm, and used the RPA to calculate the modification in phonon frequencies coming from electron-electron interaction [33]. Their result, one later realized, agreed with that obtained by Bardeen in his 1936 paper in which he used mean-field theory to take electron-electron interactions into account [34]. However, Staver's further progress was tragically cut short when he died in a skiing accident not long after the publication of his paper.

As I started work on the problem, I found it easy to reproduce the Bardeen-Bohm-Staver results, but as I carried out the various canonical transformations that were analogous to those used for electron interactions only, I found my results were not self-consistent. One morning in 1954, as I described my approach to John Bardeen [I was his postdoc and had a desk in a corner of his office], he noticed that I had not been consistent in including the phonon contribution to the density fluctuation coordinates I introduced to describe the coupled collective modes. Once I did this, everything fell into place, so I suggested we write a joint paper describing the results. We calculated the longitudinal sound velocity of simple metals and found good agreement with experiment. Importantly, we found that when full account was taken of the screening of electron-electron and electron-ion interactions, the screened attractive phonon-induced interaction between electrons was in general comparable to their repulsive screened Coulomb interaction. For quasi-electrons lying close to the Fermi surface, at low frequencies their screened frequency-dependent attractive phonon-induced interaction could win out over their screened Coulomb repulsion, and one gets a screened attractive interaction of the form proposed by Frohlich.

As Nozieres and I showed subsequently [28, 29] the result Bardeen and I had obtained can be written in especially simple fashion for a quantum plasma,

$$V_{eff}(q,\omega) = [4\pi e^2/q^2 \varepsilon(q,0)] [1 + \omega_q^2/(\omega^2 - \omega_q^2)] = [4\pi e^2/q^2 \varepsilon(q,0)][\omega^2/(\omega^2 - \omega_q^2)] \quad [12]$$

where $\omega_q$ is the frequency of the phonon being exchanged, and the second term on the rhs of Eq. [12] is the phonon-induced interaction. At zero frequency, it just cancels the repulsive screened Coulomb interaction, so for jellium one is left with a frequency dependent interaction which is attractive for frequencies less than a typical phonon frequency, $\omega_q$.

Bardeen and I concluded in our subsequent paper [6] that since full account had now been taken of the Coulomb interaction, our renormalized interaction should provide a good starting point for the development of a microscopic theory of superconductivity. This proved to be the case, when BCS [7] emerged just two years later.

I first presented our results at the 1954 Solvay Congress. In doing so I noted that our work demonstrated that Heisenberg's attempts to get superconductivity solely from electron-electron interactions was doomed to failure. When I finished, Pauli made the first remark: "I always told that fool Heisenberg that he was wrong", a remark that understandably did not make it into the printed version of the discussion. Our prediction that our interaction could form the basis for a microscopic theory of superconductivity is also not in the printed version, because I had to submit my report to the Congress some months in advance, before Bardeen and I had finished work on our paper.

**BCS Emerges**

In 1956, Leon Cooper, who was my successor as a post-doc with John Bardeen, proposed a simple toy model for the Bardeen/Pines/Frohlich interaction, in which the quasiparticle interaction was constant for electrons lying within a characteristic phonon energy of the Fermi surface, and zero otherwise, and showed that it gave rise to a bound state near the Fermi surface [more correctly, it gives rise to an instability of a normal Fermi liquid against pair creation] [35].

In early 1957, John Bardeen's graduate student, Bob Schrieffer, while riding on a New York subway, came up with a candidate wave function for the ground state of a superconductor, one in which pairs of electrons condense into a single quantum state. Schrieffer's wave function formed the basis of the microscopic theory of superconductivity, known subsequently as BCS, that was first submitted for publication a few weeks later by Bardeen, Cooper, and Schrieffer [7]. Their paper solved, at long last, what had been the major challenge in correlated electron systems, and indeed in all of theoretical physics, for some 46 years.

That spring, while lecturing on BCS to my class at Princeton, I developed a simple extension of the Bardeen/Pines/Frohlich interaction to polyvalent materials that enabled me to show that it could explain the appearance of superconductivity in the periodic table [36]. Jellium did not superconduct, but when one allowed for the Umklappprocesses present because of the ionic lattice structure, the phonon-induced interaction would bring about superconductivity. It provided a microscopic justification for the Matthias rules for finding superconductors [37], and led to the successful prediction that superconductivity would be found in Mo, W, Y, Sc, and Pd.

**Plasmons and Plasmonics**

While in the years immediately after the pioneering experiments of Ruthemann and Lang, electron energy loss experiments showed that in many solids the measured energy loss continued to be not far from the calculated free valence electron plasma quantum, the view that one was observing a new collective mode was not universally accepted. Ladislaw Marton, who was responsible for a number of these experiments, organized a special session at the 1955 Spring Meeting of the American Physical Society to discuss the issue; he invited MIT's John Slater, who was at the time arguably the world leader in calculating electronic band structures in solids, to present his view that the measured characteristic energy losses represented inter-band transitions, while I was invited to make the argument that these represented quantized plasma oscillations. After we both spoke, Marton invited the audience to vote on which interpretation they found convincing. Although I thought I had settled the issue by invoking Mott's sum rule arguments that explained the general agreement of the valence electron plasma quantum with experiment, I lost the popular vote by about a two to one margin!

However, further experiments soon proved me right. Electron energy loss experiments on a number of other materials showed that in over 50 solids, including a number of semiconductors and insulators, the measured energy loss was close to the calculated free valence electron plasma quantum [8]. Moreover, the angular distribution of energy losses in Al was measured by Watanabe [38] and shown by Ferrell to reflect the dispersion of the quantized valence electron plasma oscillation [39]. In an invited talk I gave in the spring of 1956, I therefore felt justified in naming their quanta as plasmons, and making the argument that plasmons should join phonons and magnons as leading members of the family of collective elementary excitations in solids [8] [28]. Soon afterwards, Ritchie [40] called attention to the possibility of surface plasmons associated with a wave of surface charge bound at the vacuum–sample interface and his simple result for their energy, $[\hbar\omega_p/2^{1/2}]$, explained a set of additional energy losses that been present in a number of experiments.

Sixty years later, plasmons continue to be a subject of considerable research interest, and electron energy loss spectroscopy continues to be an important tool to study their behavior, as discussed in a recent review by Roth et al [41]. Somewhat unexpectedly, plasmons have turned out to be not only of continuing interest in fundamental research, but of very considerable practical importance, as *plasmonics*, the study of devices based on controlling surface plasmons, has emerged in the early years of this century as a major subfield of nanoscience [9]. I have learned from the web that there have been five Gordon Research conferences on Plasmonics and a sixth is scheduled for 2016. There is a Springer Journal, *Plasmonics*, and the editors of a recent book-length collection of review articles [42] argue that plasmonics has entered the curriculum of many universities as a stand alone subject, or as part of a course on nano-electronics. One of the Royal Society of Chemistry's 2015 Faraday Discussions was devoted to Nanoplasmonics, and in his introductory lecture there, Mark Brongersma noted that there are "over 10 000 publications every year on the topic of plasmonics and the number of publications has been doubling about every three years since 1990" [43].

A large body of theoretical and experimental work on plasmons in graphene [44] shows that new materials continue to offer a fertile ground for their study and application. For example, Fei et al have carried out infra-red nano-imaging experiments that establish graphene/SiO2/Si structures as a tunable plasmonic medium. They conclude that "graphene may be an ideal medium for active infrared plasmonics" [45].

**The Helium Liquids**

The emergence of "The Many Body Problem" as a significant sub-field of theoretical physics in the 1950's was based on a simple premise—"have organizing principle, will travel"---so those of us who worked on electron liquids thought about applying what we had learned to their sister quantum liquids, the helium and nuclear liquids, and vice versa. Thus in our early papers, Bohm and I had noted that the RPA could easily be extended to neutral liquids, and we argued that if the restoring force were strong enough, one should find collective collisionless modes in these materials as well. We had the excitation spectrum of superfluid $^4$He in mind, but a straightforward application of the RPA to liquid $^4$He or its fermion sister, $^3$He, does not work because the bare interaction between the He atoms is much too strong; one needs a way to determine how, in these liquids, the strong short range correlations between the helium atoms modifies their bare interaction and determines the restoring forces responsible for their collective modes.

Landau did this for liquid $^3$He in the limit of very long wavelengths, by introducing moments of his phenomenological effective interaction between quasiparticles on the Fermi surface; he predicted zero sound, the neutral analogue of plasma oscillations, but offered no prescription for extending his theory to finite wavevectors. Polarization potential theory [10] offers a *phenomenological* way to do this, and I now describe it in some detail because it provides an excellent example of the power of phenomenology to uncover the organizing principles at work in quantum liquids and, when applied to the helium liquids, determine the effective quasiparticle interactions that lead to the emergence and subsequent interactions of the elementary excitations found in liquid $^4$He and $^3$He [46], and in $^3$He-$^4$He [47] mixtures.

The initial impetus for the theory came from experiment—the INS experiments on liquid $^4$He by David Woods [48] on the variation with temperature of the energy and lifetime of excitations of energy ~7.4K that are found at wavevectors ­0.38 A; he found no appreciable change in their energy and only a modest increase in their lifetime on going from temperatures well below the superfluid transition temperature to those substantially above it. These results led me to propose that the phonon-maxon-roton spectrum in superfluid $^4$He must be a collisionless mode that is the neutral analogue of plasma oscillations and is brought about by an effective scalar polarization potential,

$$\varphi_s[q,\omega] = f_q^s <\rho[q,\omega] \quad [13]$$

whose strength, $f_q^s$, is sufficient to place the excitations above those characteristic of single-particle excitations in the normal state [49]. It seemed natural to expect that the polarization potentials for $^3$He and $^4$He would be similar, so I predicted that the zero sound mode in $^3$He would continue to exist at wavevectors and temperatures for which Landau's Fermi liquid theory was no longer valid, disappearing only when it became damped by decay into particle-hole excitations, a prediction that was eventually verified in INS experiments on $^3$He by Skold et al [50].

In the further development of polarization potential theory a decade later our research group in Urbana took into account a significant physical effect whose importance was first stressed by Feynman and Cohen [51], the influence of backflow on the excitation spectra. Thus, as a particle moves in the liquid, in addition to inducing density fluctuations, it induces current fluctuations, which act back on it, thereby changing its effective mass. For liquid $^4$He, the induced current fluctuations, $<\mathbf{j}(q,\omega)>$, give rise to a vector field,

$$\mathbf{A}_{pol}[q,\omega)] = f_q^v <\mathbf{j}(q,\omega)> \quad [14],$$

that couples to the current fluctuations, $\mathbf{j}_q$, through $\mathbf{j}_q \cdot \mathbf{A}_{pol}(q,\omega)$. Since backflow is a longitudinal phenomenon, particle conservation links the induced current and density fluctuations according to

$$<\mathbf{q}\cdot\mathbf{j}(q,\omega)> = -i\omega <\rho(q,\omega)> \quad [15],$$

so including backflow adds a frequency-dependent restoring force for collective modes, and the new polarization potential is given by

$$\varphi_{pol}(q,\omega) = [f_q^s + (\omega^2/q^2) f_q^v] <\rho(q,\omega)> \quad [16].$$

In working out the consequences of Eq. [16], it is useful to introduce the density-density response function, $\chi[q,\omega]$, whose imaginary part is measured in INS experiments [28]; it can be written as

$$\chi[q,\omega] = \chi_{sc}(q,\omega)/[1-[f_q^s + (\omega^2/q^2) f_q^v]\chi_{sc}(q,\omega) \quad [17]$$

where $\chi_{sc}(q,\omega)$ is the response of the density fluctuations to an external field plus the polarization potential, Eq. [16]. Determining the excitation spectrum then involves developing models for $f_q^s$, $f_q^v$, and $\chi_{sc}(q,\omega)$.

The parameter, $f_q^s$, is the Fourier transform of a configuration space pseudopotential, $f[r]$, that describes the effective interaction responsible for collective modes in the density fluctuation spectrum. In building a simple physical model for $f[r]$, Aldrich and Pines [45], hereafter AP, argued that in the liquid the strong repulsive interaction between the bare He atoms would prevent them from sampling too much of it, while the longer range attraction would be comparatively unchanged from its value for a gas. A simple model interaction for f[r] in $^4$He that incorporates these ideas is a soft-core repulsive interaction,

$$f[r] = a[1-r/r_c]^8 \quad r<r_c \quad [18],$$

that is joined to the attractive long range part [assumed to be identical to that for bare atoms] by a simple fitting function. Because $f_0^s$, the spatial average of f[r], is known from the compressibility, for a given choice of range, $r_c$, the strength, a, of the soft core repulsion is uniquely determined, while $r_c$ is determined by the pressure-dependent zero point motion of the atoms in the liquid. $^4$He. For $^3$He one expects the range of the repulsion to increase because of its increased zero point motion, while the Pauli principle will lead to a further increase in range for particles of parallel spin; one can then use the spin susceptibility as well as the compressibility to fix its strength and that for particles of antiparallel spin [46], [53].

For $^4$He the strength of the backflow potential, Eq.(14), determines the momentum-dependent quasiparticle effective mass, $m_q^*$,

$$m_q^* = m + N f_q^v \quad (19)$$

and AP argue that its fall-off in momentum space from $m_0$ will roughly resemble that of $f_q^s$ since the induced current and density fluctuations should exhibit similar behavior.

For $^3$He one needs separate polarization potentials for parallel and antiparallel spin quasiparticles; these reduce, in the longwave length limit, to the self-consistent fields introduced by Leggett in his formulation of Landau Fermi liquid theory [52]; $m_q^{\bullet}$ is the average single quasiparticle–quasihole effective mass that, in the long wavelength limit, becomes the quasiparticle mass, m*,

$$m^* = m + N[0]f_0^v = m\,[1+F_1^s/3] \quad [20]$$

Before applying these pseudopotentials, there is one more physical effect that needs to be taken into account, mode-mode coupling to multiparticle excitations for $^4$He, and to multipair excitations for $^3$He. This can be done through simple models for the screened response function, $\chi_{sc}$ in Eq.[17]. One arrives at a unified and quantitative account of the density fluctuation excitation spectrum at atmospheric pressure in $^3$He [46] and its changes at higher pressures [54]; the pressure dependence of the phonon-maxon roton spectrum in $^4$He [45]; and the anomalous phonon dispersion found there [55]. I refer the interested reader to the original papers and the two review articles [10] for the details.

Independent support for the AP polarization potentials for $^4$He came from the variational calculations of the phonon-maxon-roton spectrum by Manousakis and Pandharipande [56], who, when they included backflow, but neglect mode-mode coupling to multiparticle excitations, obtained results identical to the AP results [using this same approximation] for wavevectors up to 1.5 A.

$^3$He-$^4$He mixtures represent an interesting quantum liquid, because the introduction of $^3$He atoms into $^4$He reduces the system density and so changes the restoring forces responsible for the $^4$He collective modes. Hsu and I explored these changes [57]; our pseudopotentials were then used to obtain a quantitative account of the elementary excitation spectrum measured in INS experiments on the mixtures [47]. It turns out that mode-mode coupling effects are important and the results are sufficiently sensitive to the form of the $^3$He quasiparticle spectrum used to fit specific heat experiments that one can distinguish between different proposals for this quantity and establish a link between INS and specific heat experiments on the mixtures.

The AP pseudopotentials led to a detailed description of rotons and their interactions in $^4$He. Contrary to Feynman's poetic picture of the roton "as the ghost of a vanishing vortex ring", at svp a roton is a quasiparticle of mass ~2.8$m_0$ moving in an average attractive self-consistent field ~2K produced by the other rotons [45]. Bedell et al [58] built on this result to develop roton liquid theory, a Bose liquid analogue of Landau's Fermi liquid theory; it takes into account the influence on a given roton of the average self-consistent field produced by the other rotons and enabled them to describe the temperature variation of the specific heat and superfluid density in terms of a few phenomenological parameters.

An Urbana group, hereafter BPZ [59], then developed a configuration space pseudopotential for roton-roton interactions that is based on the concept of rotons as quasiparticles interacting via a pseudopotential of the form, Eq. [20]; the range of their soft-core repulsion was found to be somewhat larger, varying from 3.4A at svp to ~ 3A at 25 bar, with its strength determined by the parameters given in roton liquid theory. BPZ constructed scattering amplitudes from their pseudopotentials and used these to calculate roton liquid parameters, two-roton bound states, and roton lifetimes, with results in excellent agreement with Raman scattering [60] and other experiments.

Additional support for the phenomenological pseudopotentials proposed for $^3$He came when these were used to calculate the scattering amplitudes for quasiparticles near the Fermi surface and obtain a quantitative account of its transport properties [61]. When the average attractive interaction in the p-wave channel was then calculated by Bedell and the author, we found the pressure variation of the superfluid transition temperature of He, and the BCS strong coupling corrections agreed surprisingly well agreement with experiments on its superfluid behavior [62]. Overall, the agreement between theory and experiment using some nine distinct angular averages of the BP scattering amplitude led us to conclude that it must be very nearly the correct one.

An interesting recent development is the experimental discovery of a large momentum transfer roton collective mode in two-dimensional $^3$He by Godfrin et al [63].

For any given wavevector, the particle–hole pairs that can damp such a mode possess a maximum energy, so if the roton collective mode at this wavevector is larger, damping will be minimal, and the mode can be observed, and this is what Godfrin et al established to be the case for a considerable range of wavevectors.

**From Zero Sound in Nuclei to Superfluidity in Neutron Stars**

I grew up scientifically before our current age of specialization. Being a theoretical physicist meant following all of physics, so when I learned about the giant dipole resonance in nuclei in 1951, I decided to explore whether it could be a collective motion of neutrons against protons that was the nuclear analogue of plasma oscillations. I gave this as a thesis problem to my first [and only] Penn graduate student, Mel Ferentz. After I moved to Urbana, as we were beginning to write up the RPA results that supported this proposal we discussed the issue of how best to go from nuclear matter to a finite nucleus with Murray Gell-Mann, whose PhD concerned nuclear theory, and who was on a visit from Chicago in the summer of 1953, Murray came up with a solution, and we published our results as a PRL [11]. This may have been the first application to nuclear physics of concepts developed in condensed matter.

A second opportunity to apply these concepts arose in Copenhagen in June, 1957. In the course of a lecture on the emerging [and not yet fully described] BCS theory to an audience at the Niels Bohr Institute, I suggested that
since the key to superconductivity was an attractive interaction between fermions, perhaps nuclei would display superfluid behavior. Aage Bohr and Ben Mottelson, who were in the audience, liked the idea and we found over the summer that it offered promise of explaining pairing phenomena in nuclei. I introduced the concept of nuclear superconductivity to the audience at an international conference on nuclear physics in Rehovoth that fall, and our joint paper on the topic appeared the following year [12].

Arkady Migdal, who had developed a Fermi liquid description of nuclei, subsequently suggested that if indeed neutron stars existed, it was likely these would contain abundant quantities of nuclear superfluids [64]. The 1967 discovery of pulsars turned neutron stars from a theorist's dream into reality and soon afterwards the observation of their behavior following a glitch in their rotational period provided strong evidence for superfluid behavior by the nuclear matter inside neutron stars [65]. There followed, a decade later, the application of polarization potentials to nuclear matter that enabled our group in Urbana to determine phenomenologically the effective quasiparticle interactions in neutron stars and use these to calculate the superfluid transition temperatures and energy gaps for the cosmic superfluids in the crust and core of pulsars that we now recognize as the most abundant superfluids found in the universe [13].

**The Random Phase Approximation**

In retrospect, the RPA marked a turning point in our understanding of interacting many-body systems: for electron liquids it showed how electron interactions lead to those two hallmarks of their emergent behavior: plasma oscillations and screening and to the concept of weakly interacting quasi-electrons [electrons plus their co-moving screening clouds]. When used only for a limited wavevector regime in the collective description of electrons in metals, it made possible a successful microscopic calculation of the correlation energy and quasiparticle properties of the alkali metals.

For a general many-body system, it showed how particle interactions can give rise to collisionless collective modes while offering a systematic way to go beyond the Hartree-Fock approximation in calculating system properties. Sixty-plus years later, the RPA continues to play a significant role in nuclear physics [66], bosonic field-theory [67], the quark-gluon plasma [68], many-fermion solvable models [69], and especially in computational chemistry and materials science. A recent review by Ren et al [70], to which the interested reader is referred, describes the impact of the RPA in the theoretical chemistry and materials science community, cites some thirty articles that indicate the renewed and widespread interest in the RPA during the period 2001-2011, discusses how it enables one to derive the $1/r^6$ interaction between spatially separated closed shell electron systems, and, shows, in some detail, how the RPA enables one to go beyond density functional theory in computing ground state energies.

**The paradigm shift for SCES**

A major paradigm shift in theoretical physics took place during the years in which SCES emerged as a distinct sub-field of condensed matter theory. When I began work on my PhD., and for some years afterwards, the reigning paradigm was reductionism, with its top-down approach that began by identifying the interactions between the basic constituents of matter and then seeking to solve a Hamiltonian that described these. Today most theorists follow instead an "emergent" paradigm, using experiment to identify emergent collective behavior and exploiting phenomenology to identify candidate organizing principles that might explain it, before attempting to devise and solve a model that incorporates those organizing principles.

Looking back, the papers David Bohm and I wrote reflect a transition between these two modes of thinking. We began with a microscopic Hamiltonian, but we looked at it in a different way, focusing on the way it described an interaction between collective variables, the density fluctuations, whose time dependent behavior might provide the key to understanding electron liquids. In so doing we arrived at a new elementary excitation in solids, plasmons, the quantized collective modes of the valence electron plasma, and a new physical picture of a metal as made up of three interacting elementary excitations, plasmons, phonons, and quasi-electrons, and argued that the comparatively weak screened interaction between the quasi-electrons could typically be dealt with at a microscopic level.

The helium and nuclear quantum liquids resemble electron liquids in a number of ways, but it has proven difficult to develop a first-principles analytic account of their emergent behavior. We do however possess variational calculations [51,56] and the phenomenological approach described above: polarization potentials that include backflow, which, together with mode-mode coupling, incorporate candidate organizing principles that appear capable of providing us with a physical picture of what determines their elementary excitations, transport properties, and superfluid transition temperatures. Their application to the entire family of helium liquids at many different pressures have led to results that agree remarkably well with experiment. Thus while we do not yet have a complete microscopic derivation of the effective quasiparticle interactions that play a central role in this phenomenological approach, given the extent to which these lead to agreement with experiment, phenomenology may suffice.

What are the respective roles of phenomenology and microscopic models in current research on SCES?  Understanding Kondo lattice materials has been a major challenge since the discovery of heavy electrons in these materials some forty years ago [71]. Doing so has turned out to require an emergent perspective, as it is only through a phenomenological two fluid description that it has finally proved possible to disentangle heavy electron and local moment behavior, and arrive at the conclusion that the emergence and subsequent behavior of heavy electrons is a *collective* phenomenon brought about by the *collective* hybridization of local moments and background conduction electrons [72]. As discussed by Lonzarich et al, we can now understand why all previous attempts to develop microscopic theories of Kondo lattice behavior have failed, and we are hopefully now on the threshold of developing microscopic theories that incorporate this key organizing principle [73].

For some three decades now, understanding the emergent unconventional superconductivity found in the heavy electron, organic, cuprate, and iron-based materials has been the central problem in our field. Our theoretical community has responded with a number of different candidate organizing principles/emergent approaches to achieve this understanding. Their common features are that the candidate mechanisms for their unconventional superconductivity are purely electronic, rather than the phonon-induced interaction between electrons responsible for conventional superconductivity,and that their origin is closely tied to the fact that in all of these materials superconductivity emerges on the border of antiferromagnetism. [3].

Two classes of experiment-driven theories emerged within the year following the discovery of the high $T_c$ cuprate superconductors: the use of a Hubbard model to explore the possibility that the mechanism was a spin-fluctuation induced interaction between the quasiparticles that led to d-wave pairing [74], and the proposal that the materials are best described in terms of a resonating valence bond between the Cu atoms [75]. A few years later, groups in Tokyo [76] and Urbana [77] focused on the quantum critical spin fluctuations seen in NMR and INS experiments that reflect their close approach to antiferromagnetism and argued that these provide the glue for their unconventional superconductivity and produce an effective interaction that differs from that used in the Hubbard or RVB models. Their strong coupling calcul;ations of  the effectiveness of the quantum critical spin-fluctuation interaction between quasiparticles went well beyond the earlier Hubbard approach and yielded d-wave pairing and superconducting transition temperatures of 100K or larger [78].

There is in addition a large body of theoretical work based on using their "Mottness", the localization brought about by the strong particle interactions as key to their behavior [79], while a very considerable effort continues to be expended on pursuing the consequences of an RVB description [80] and finding better numerical solutions for a single band Hubbard model Hamiltonian that its proponents believe contains the key elements needed to understand the cuprates [81]

As of this writing there is increasing experimental evidence that all four families of unconventional superconductors exhibit, over some regime of doping. pressure, and temperature, both localized and itinerant behavior and possess a quantum critical point that marks the end of localized behavior, but no general consensus has emerged on which of these candidate organizing principles and related models plays the central role in determining their remarkable normal and superconducting states. It is to be hoped that an experiment-based consensus will emerge during the coming fourth decade of theoretical work on the problem.

In concluding let me note that I have not considered one considerable body of work in condensed matter theory, computational approaches that are based on using sophisticated algorithms and large scale computers to carry out, in some cases, ab initio calculations. Recent examples include DMFT theory [82] and the just-cited studies of the Hubbard Hamiltonian [81]. While these have led to a "proof of concept" of the effectiveness of frequency-dependent spin-fluctuation-induced superconductivity in the cuprates [78]) and explained emergent behavior (eg the temperature-dependent heavy electron density of states in CoIrIn5 obtained using DMFT [823), the path from the starting Hamiltonian to the finished product is not always clear, nor is the extent to which one will be able to use these in the future to establish the physical origin of the emergent behaviors we seek to explain [2].

**Acknowledgements**

As the alert reader may have noticed, in writing this perspective I have not hesitated to use materials in some of my earlier reviews and books. I should like to take this opportunity to thank those with whom I collaborated or discussed the papers cited here for their invaluable role in developing and applying the concepts these describe: my students, Chuck Aldrich, Victor Barzykin, Mel Ferentz, Daryl Hess, Wei-chan Hsu, Setsuo Ichimaru, Naoki Iwamoto, Philippe Monthoux, and Philippe Nozieres; my postdocs,Tom Ainsworth, Ali Alpar, Sasha Balatsky, Kevin Bedell, Andrey Chubukov, Tony Leggett, Dirk Morr, Khandker Quader, Joerg Schmalian, Jacob Shaham, Branko Stojkovic, and Yi-feng Yang; and my colleagues, Elihu Abrahams, Gordon Baym, Aage Bohr, David Campbell, Piers Coleman, Nick Curro, Zachary Fisk, Richard Ferrell, Igor Fomin, Hans Frauenfelder,Jacques Friedel, Murray Gell-Mann, Lev Gor'kov, Laura Greene, John Hubbard, Miles Klein,Gabi Kotliar, Ekhard Krotscheck, Bob Laughlin, Gil Lonzarich, Stratos Manousakis, Neville Mott, Ben Mottelson, Satoru Nakatsuji, Vijay Pandharipande, Chris Pethick, Myron Salamon, Bob Schrieffer, Charlie Slichter, Joe

Thompson, Jochem Wambach, Peter Wolynes, and Fred Zawadowski. I dedicate this "remembrance of things past' to them, to my mentors in condensed matter theory, John Bardeen, David Bohm, and Conyers Herring, and especially to Suzy, whose unwavering support throughout my career has been more valuable than words can ever express.